\documentclass[12pt]{article}
\usepackage{latexsym}
\usepackage{amsmath,,calrsfs}
\usepackage{amsfonts}
\usepackage{amssymb}
\usepackage{amscd}
\usepackage{bbm}
\usepackage{fancybox}
\usepackage{cite}
\usepackage{amsmath,amsfonts,amsbsy}
\usepackage{pstricks,pst-node}
\usepackage[small,bf,hang]{caption2}
\usepackage{graphicx}
\usepackage{epsfig}
\usepackage{psfrag}
\usepackage{comment}

\usepackage{float}

\psset{unit=1.3cm,linewidth=.5pt,radius=.2}  % controls the size of diagrams

\usepackage{multirow}                     % tables cells spanning multiple rows
\usepackage{float}                          % to force floaters to stay Here
\usepackage{lscape}                         % landscape laid out pages
\usepackage{bm}

\newcommand{\beq}{\begin{equation}}
\newcommand{\eeq}{\end{equation}}
\newcommand{\bi}{\begin{itemize}}
\newcommand{\ei}{\end{itemize}}
\newcommand{\bt}{\begin{tabular}}
\newcommand{\et}{\end{tabular}}
\newcommand{\bc}{\begin{center}}
\newcommand{\ec}{\end{center}}

\newcommand{\be}{\begin{equation}}
\newcommand{\ee}{\end{equation}}
\newcommand{\bea}{\begin{eqnarray}}
\newcommand{\eea}{\end{eqnarray}}
\newcommand{\ba}{\begin{array}}
\newcommand{\ea}{\end{array}}

\def\bbox{{\,\lower0.9pt\vbox{\hrule \hbox{\vrule height 0.2 cm
\hskip 0.2 cm \vrule height 0.2 cm}\hrule}\,}}
\newcommand{\dsl}{\pa \kern-0.5em /}

\makeatletter \@addtoreset{equation}{section} \makeatother

\def\slashchar#1{\setbox0=\hbox{$#1$}           % set a box for #1
   \dimen0=\wd0                                 % and get its size
   \setbox1=\hbox{/} \dimen1=\wd1               % get size of /
   \ifdim\dimen0>\dimen1                        % #1 is bigger
      \rlap{\hbox to \dimen0{\hfil/\hfil}}      % so center / in box
      #1                                        % and print #1
   \else                                        % / is bigger
      \rlap{\hbox to \dimen1{\hfil$#1$\hfil}}   % so center #1
      /                                         % and print /
   \fi}

\begin{document}

\begin{titlepage}%1
\begin{center}

\hfill  DAMTP-2014-37

\vskip 1.5cm

{\Large \bf  Matter coupling in 3D ``Minimal Massive Gravity''}\\
%{\Large \bf and its asymptotically de Sitter solutions}

\vskip 1cm

{\bf Alex S. Arvanitakis, Alasdair J. Routh and 
Paul K.~Townsend} \\

\vskip 1cm

Department of Applied Mathematics and Theoretical Physics,\\ Centre for Mathematical Sciences, University of Cambridge,\\
Wilberforce Road, Cambridge, CB3 0WA, U.K.\

\vskip 5pt

{email: {\tt A.S.Arvanitakis@damtp.cam.ac.uk, A.J.Routh@damtp.cam.ac.uk, P.K.Townsend@damtp.cam.ac.uk}} \\

\end{center}

\vskip 0.5cm

\begin{center} {\bf ABSTRACT}\\[3ex]
\end{center}

The ``Minimal Massive Gravity''  (MMG) model  of massive gravity in three spacetime dimensions (which has the same anti-de Sitter (AdS) bulk properties
as ``Topologically Massive Gravity'' but improved boundary properties) is coupled to matter.  Consistency requires a particular matter  source tensor, which is quadratic in the stress tensor.  The consequences are explored for an ideal fluid in the context of asymptotically de-Sitter (dS) cosmological solutions, which bounce smoothly from contraction to expansion. Various vacuum solutions are also found, including  warped (A)dS, and (for special values of parameters) static black holes and an (A)dS$_2\times S^1$  vacuum.

\end{titlepage}

\newpage
\setcounter{page}{1} 
%\tableofcontents

\newpage

%%%%%%%%%%%%%%%%%%%%%%%%%%%
\section{Introduction}

The three-dimensional (3D) massive gravity model known as ``Topologically Massive Gravity'' (TMG)  \cite{Deser:1981wh} has attracted much attention over the last decade, 
particularly in the context of asymptotically anti-de Sitter (AdS) boundary conditions for which there is the possibility of a consistent quantum completion 
via a holographically dual  conformal field theory.  However, using the methods initiated  in the context of  3D Einstein gravity by Brown and Henneaux \cite{Brown:1986nw},
one finds that one of the two (left/right) central charges  (which differ because TMG violates parity \cite{Kraus:2005zm}) must be negative whenever the bulk spin-2 mode has 
positive energy, i.e. is not a ``ghost''; as reviewed in \cite{Kraus:2006wn}.

An alternative to TMG that circumvents this problem was  recently proposed \cite{Bergshoeff:2014pca}. It has identical bulk properties to TMG
but boundary central charges that are both positive within a range of the parameter space. This model was called ``minimal massive gravity'' (MMG), partly
because it shares with TMG the feature of propagating a single spin-2 mode, with no other local degrees of freedom, and partly because its equation of motion
is such that  coupling to matter appears impossible. The latter point can be understood by inspection of the MMG equation, which is 
\begin{equation}\label{sourcefree}
E_{\mu\nu} \equiv \bar\Lambda_0\,  g_{\mu\nu}  + \bar\sigma G_{\mu\nu} + \frac{1}{\mu} C_{\mu\nu} + \frac{\gamma}{\mu^2} J_{\mu\nu} =0\, , 
\end{equation}
where $\bar\sigma$ and $\gamma$ are dimensionless constants,  $\bar\Lambda_0$ is a cosmological parameter (with dimensions of mass-squared) and $\mu$ is the TMG mass parameter; we use the 
notation of \cite{Bergshoeff:2014pca} for ease of comparison.  When $\gamma=0$ this equation is the (cosmological) TMG equation, with $G$ the Einstein tensor and $C$ the Cotton tensor; the new feature of MMG is the $J$ tensor, which is\footnote{This differs by a sign from the definition of this tensor in  the published version of \cite{Bergshoeff:2014pca} (and v3 on the arXiv) because there is a mismatch there between the dreibein and tensor formulations; the results of  \cite{Bergshoeff:2014pca} that relate the two, which we shall use later,  are valid for the definition of $J$ given here.}
\begin{equation}
J^{\mu\nu} = \frac{1}{2\det g} \, \varepsilon^{\mu\rho\sigma}\varepsilon^{\nu\tau\eta} S_{\rho\tau}S_{\sigma\eta}\, , 
\end{equation}
where $S_{\mu\nu}$ is the 3D Schouten tensor.  The novelty of this particular curvature-squared extension of TMG is that the $J$-tensor does {\it not} satisfy a Bianchi-type identity; in fact, 
\begin{equation}\label{notB}
\sqrt{-\det g}\, D_\mu J^{\mu\nu} =    \varepsilon^{\nu\rho\sigma} S_\rho{}^\tau C_{\sigma\tau}\, .
\end{equation}
This means that  the MMG equation is {\it not the Euler-Lagrange equation for any action constructed from the metric and its curvature tensor}. In spite of this,  the MMG equation 
is consistent because it  implies that the right hand side of  (\ref{notB}) is zero; in other words, the $J$ tensor may be viewed as a kind of stress tensor for the metric 
that is conserved by virtue of the metric equation, even when that equation includes $J$ itself. 

Following \cite{Bergshoeff:2014pca},  let us now consider how MMG might be coupled to matter, represented by some stress tensor $T$ satisfying the usual conservation condition
\begin{equation}\label{Tcon}
D^\mu T_{\mu\nu}=0\, 
\end{equation}
as a consequence of the matter equations of motion.  
For TMG we only have to modify the source-free equation to $E_{\mu\nu}=T_{\mu\nu}$ (for a particular choice of units for the 3D Newton constant), but this equation is inconsistent for MMG; it  implies, because of (\ref{notB}), that  $\varepsilon^{\nu\rho\sigma} S_\rho{}^\tau T_{\tau\sigma} =0$. This is an unacceptably strong constraint  because it requires the stress tensor to be  proportional to the Einstein tensor, which implies (in 3D) the absence of any propagating  modes. 

Although the standard matter coupling is not available when $\gamma\ne0$, we could consider  an equation of the form
\begin{equation}\label{sourced}
E_{\mu\nu} =  {\cal T}_{\mu\nu}\, , \qquad {\cal T}_{\mu\nu} = T_{\mu\nu} + {\cal O}\left(\gamma\right)\, , 
\end{equation}
in which case consistency requires the  source tensor ${\cal T}$ to satisfy, as a consequence of (\ref{Tcon}) and hence of the matter equations of motion,
\begin{equation}\label{identity}
D_\mu {\cal T}^{\mu\nu} = \frac{\gamma}{\mu\sqrt{-\det g}}\,  \varepsilon^{\nu\rho\sigma} S_\rho{}^\tau {\cal T}_{\sigma\tau}\,  . 
\end{equation}
The main result of this paper is a construction of such a source tensor for MMG.  This construction leads directly  to a simpler parametrization  in which the cosmological parameter $\bar\Lambda_0$ is replaced by the constant energy density of a ``dark energy'' source, and  in this parametrization we are able to express the matter source tensor directly  in terms of the source-free parameters. 

We explore the consequences of this result  in the simple context of  FLRW cosmology, i.e. time-dependent  but homogeneous and isotropic solutions with an ideal fluid source of energy density $\rho$ and pressure $p$, for which 
\begin{equation}
T_{\mu\nu} = \left(\rho + p\right) u_\mu u_\nu  + p\, g_{\mu\nu}\, , 
\end{equation}
where $u_\mu$ is the fluid 3-velocity.  Assuming a linear equation of state, we find that MMG leads to the same Friedmann equation as TMG but with the energy density of the fluid  replaced by a $\gamma$-dependent ``effective energy density''  $\rho_{\rm eff}(t)$.  Focusing on the case of flat universes, we show that the big-bang singularity of the TMG case is replaced by a smooth ``bounce'', so that the complete spacetime is asymptotic to the same de Sitter (dS) spacetime in both the far past and the far future. This appears to be a further illustration, now in the context of dS vacua,  of how MMG has better short distance behaviour than TMG. 

In some respects, MMG  is similar to the parity-preserving ``New Massive Gravity'' (NMG) model of  massive 3D gravity \cite{Bergshoeff:2009hq}. For example,  the trace of the $J$-tensor is the same (up to normalisation) as the trace of the curvature-squared $K$-tensor of the NMG equation, which is proportional to the curvature-squared scalar $K$ that appears in the NMG action.  It is not clear to us whether this remarkable coincidence has any significant implications, but the fact that both models involve curvature-squared terms  leads, in the context of maximally-symmetric vacua,  to a quadratic equation for the cosmological constant $\Lambda$ in terms of the cosmological  parameter, i.e. $\bar\Lambda_0$ for MMG.  Generically, therefore, there are two possible values of $\Lambda$ for each choice of $\bar\Lambda_0$, or none,  but for a special value of  $\bar\Lambda_0$ (which will depend on the other parameters)  there is a unique value of $\Lambda$, i.e. a unique maximally-symmetric vacuum. For NMG, special vacuum solutions become possible at this ``merger point'' (as we shall call it here); these include static black hole spacetimes that are asymptotic, but not locally isometric, to the (A)dS vacuum \cite{Bergshoeff:2009aq}. 
We show here that MMG at the merger point admits essentially the same solutions. In particular, by considering a special case, one may deduce the existence of a Kaluza-Klein  (A)dS$_2 \times S^1$ vacuum and we have verified that this is a solution of the MMG equation at (and only at) the merger point\footnote{This is again analogous to NMG, which also admits Kaluza-Klein vacua at its merger point \cite{Clement:2009gq}.}.

In addition to admitting (A)dS vacua, TMG also admits  ``warped (A)dS'' vacua   \cite{Nutku:1993eb,Moussa:2003fc,Bouchareb:2007yx,Gurses}, and it was further shown in  \cite{Anninos:2008fx}
that any model admitting such a  vacuum also admits black hole solutions that are locally isometric to it. Here we show that MMG admits warped (A)dS vacua, and also that there is a warped equivalent of the merger point for maximally-symmetric vacua. 

We shall begin with a discussion of the vacua of MMG, in particular the dS vacua and some other vacuum solutions, particularly those that become possible at the merger point. We shall then move on to matter coupling, with a derivation of the source tensor ${\cal T}$ that is needed for consistency, and we then use this result to investigate FLRW cosmologies in MMG. 

%%%%%%%%%%%%%%%%%%%%%%%%%%%%%%%%%%%%%%%%%
%%%%%%%%%%%%%%%%%%%%%%%%%%%%%%%%%%%%%%%%
\section{Some MMG vacuum solutions}

Maximally symmetric  vacua of MMG are solutions of the field equation (\ref{sourcefree}) for which $G_{\mu\nu} = - \Lambda g_{\mu\nu}$; equivalently, 
\begin{equation}
S_{\mu\nu}= \frac{1}{2} \Lambda \, g_{\mu\nu}
\end{equation}
for some cosmological constant $\Lambda$. Using this in (\ref{sourcefree}) yields the following quadratic equation for $\Lambda$: 
\begin{equation}\label{Lambdaeq}
\bar\Lambda_0 - \bar\sigma \Lambda + \frac{\gamma}{4\mu^2} \Lambda^2 =0\, . 
\end{equation}
In the TMG case ($\gamma=0$) the cosmological constant is uniquely fixed to be $\Lambda=\bar\Lambda_0/\bar\sigma$. In the MMG case ($\gamma\ne0$)  we have the two possibilities
\begin{equation}\label{mp}
\Lambda= \frac{2\mu}{\gamma} \left[\mu\bar\sigma \pm m\right] \, , \qquad m\equiv \sqrt{\mu^2\bar\sigma^2 - \gamma\bar\Lambda_0}\, . 
\end{equation}
It follows that there is no maximally-symmetric vacuum unless 
\begin{equation}
\mu^2\bar\sigma^2 - \gamma\bar\Lambda_0 \ge0\, . 
\end{equation}
This inequality is saturated when $m=0$, which we shall refer to as  the  ``merger point''; in this special case, 
there is a unique maximally-symmetric vacuum with 
\begin{equation}
\Lambda = \frac{2\mu^2\bar\sigma}{\gamma}\, , \qquad \bar\Lambda_0 = \frac{\mu^2\bar\sigma^2}{\gamma}\qquad (m=0).
\end{equation}

When $\Lambda<0$, which implies an AdS vacuum, the graviton is a tachyon at the merger point  \cite{Bergshoeff:2014pca}, but the focus of this paper will be on dS vacua. 
A modification of the linearisation results of  \cite{Bergshoeff:2014pca}, so that they apply to a dS vacuum,  leads to the conclusion that the Higuchi lower bound on the graviton mass in  dS space \cite{Higuchi:1986py} is saturated at the merger point, as is the case for NMG \cite{Bergshoeff:2009aq}.  We shall now discuss some of these solutions. 

%%%%%%%%%%%%%%%%%%%%%%%%%%%%%%%
\subsection{Black holes and Kaluza-Klein vacua}
The MMG equation (\ref{sourcefree}) at the merger point admits asymptotically (A)dS black hole solutions, analogous to those found for NMG \cite{Bergshoeff:2009aq}.  The metric take the form
\begin{equation}\label{BHmetric}
ds^2 =  \Lambda (r - r_-) (r-r_+) dt^2 - \frac{dr^2}{\Lambda (r - r_-) (r-r_+)} +r^2 d \theta^2 \,, 
\end{equation}
where $r_+\ge r_-$.  The Ricci scalar of these solutions is 
\begin{equation}
R= 6 \Lambda - \frac{2 \Lambda (r_+ +r_-)}{r}\, ,
\end{equation}
which shows that these solutions are not locally isometric to the (A)dS vacuum to which they asymptote at large $r$.  Their interpretation depends on the sign of $\Lambda$. 
\begin{itemize}

\item{$\Lambda<0$}. In this case we have an asymptotically-AdS black hole solution with an event horizon at $r=r_+$ and static exterior spacetime. 
When $r_-=r_+=r_0$  we have the zero-temperature extremal black hole solution
\begin{equation}
ds^2= - \frac{\left(r-r_0\right)^2}{\ell^2} dt^2 + \frac{\ell^2 dr^2}{\left(r-r_0\right)^2} + r^2d\theta^2\, , 
\end{equation}
where we have set $\Lambda= -1/\ell^2$.  Following \cite{Gibbons:1993sv}  we can examine the near-horizon limit by using the coordinates
\begin{equation}
\tau= \frac{\ell^2}{a} t\, , \qquad \rho= \frac{r-r_0}{a}\, , 
\end{equation}
and then taking the $a\to 0$ limit. The metric in this limit is 
\begin{equation}
ds^2 = \ell^2\left[ - \rho^2 d\tau^2 + \frac{d\rho^2}{\rho^2} \right] +r_0^2 d\theta^2\, , 
\end{equation}
which is AdS$_2 \times S^1$.

\item{$\Lambda>0$}. In this case $r=r_+$ is the cosmological horizon of a static dS vacuum, and if $r_->0$ we have a black hole in the vacuum with horizon at $r=r_-$. 
The metric is static for $r_-<r<r_+$.  This static region does not disappear in the $r_-\to r_+$ limit because the proper distance between the two horizons remains finite; instead it 
becomes a metric on  $dS_2\times S^1$. This can be seen by setting
\begin{equation}
r_\pm = r_0 \pm a
\end{equation}
and introducing the new coordinates 
\begin{equation}
\tau= \Lambda at \, , \qquad  \rho= \frac{r-r_0}{a} \, . 
\end{equation}
After taking the $a\to 0$ limit we find the metric 
\begin{equation}
ds^2 = \Lambda^{-1}\left[ - \left(1-\rho^2\right) d\tau^2 + \frac{d\rho^2}{1-\rho^2} \right] + r_0^2 d\theta^2
\end{equation}
which is a static metric on $dS_2\times S^1$; the coordinate singularities at $\rho=\pm1$ are two equal-temperature cosmological event horizons.

\end{itemize}
We have just deduced the existence of an $(A)dS_2\times S^1$  vacuum at the merger point. We have verified this directly and we have also verified that
these Kaluza-Klein vacua are not solutions of MMG away from the merger point. They are also not solutions of TMG.

%%%%%%%%%%%%%%%%%%%%%%%%%%%%%%%
\subsection{Warped (A)dS vacua}
MMG also possesses warped (A)dS vacuum solutions of all varieties. The spacelike warped AdS$_3$ metric is 
\begin{equation}\label{warp}
ds^2 = \frac{\ell^2}{(\nu^2 + 3)}\left[ -dt^2(1+r^2) + \frac{dr^2}{1+r^2} + \frac{4\nu^2}{\nu^2 + 3}\left( du + r dt \right)^2 \right] \,.
\end{equation}
where the (non-zero) warp parameter $\nu$ may be assumed  positive without loss of generality.  This metric reduces to AdS$_3$ with $\Lambda= -1/\ell^2$ when $\nu^2=1$. Other warped (A)dS vacuum metrics can be obtained from it by analytic continuation and/or (for null warping) by taking the  $(\nu\to 1)$ limit \cite{Anninos:2008fx,Anninos:2010}. 

The metric (\ref{warp}) solves the MMG equation (\ref{sourcefree}) when its parameters $\nu$ and $\ell$ are related to $(\bar\sigma,\gamma,\mu)$ by 
\begin{equation}\label{warp1}
4 \mu  \ell  \left[-6 \nu ^3+6 \nu +\bar{\Lambda}_0 \mu  \ell ^3+ \left(3-2 \nu ^2\right)
   \bar{\sigma} \mu\ell \right]+ \gamma  \left(3-2 \nu ^2\right)^2 =0\, , 
 \end{equation}
which is a deformation of equation (\ref{Lambdaeq}),  and  by
\begin{equation}
(\nu -1) \left[\gamma  \left(2 \nu ^2-3\right)-2 \mu  \ell  (3 \nu +  \bar{\sigma}\mu  \ell )\right]=0\, , 
\end{equation}
which is trivially satisfied when $\nu=1$ and otherwise equivalent to the equation 
\begin{equation}\label{warp2}
2\mu\ell \left(3\nu + \bar\sigma \mu\ell\right) = \gamma \left(2\nu^2 -3\right)\, . 
\end{equation}
Using this last equation in (\ref{warp1}) we deduce that 
\begin{equation}
3\mu\ell \left[\nu^2- \left(\frac{\bar\sigma\mu\ell}{3}\right)^2 \right] + 2\gamma\left(-\nu^3 +\nu +\frac{\bar\Lambda_0\mu\ell^3}{6}\right) =0
\end{equation}
Both this equation and (\ref{warp2}) are consistent with the TMG result that $\nu=\left|\bar\sigma\mu \ell\right|/3$, which is found on setting $\gamma=0$. 
Since warped (A)dS black holes are obtained by discrete identifications of this class of solutions \cite{Anninos:2008fx}, it follows that they also solve MMG under the same conditions.

A novel feature of MMG is that (\ref{warp1}) is a quadratic equation for $\mu\ell$ so that there are generically two values (or none) for $\mu\ell$, but there is also a ``warped merger point'' 
at which a unique value for $\mu\ell$ is possible. At this point we find that 
\begin{equation}
36\nu^2\left(1-\nu^2\right)^2 = \gamma \left(3-2\nu^2\right)^2\left[\left(3-2\nu^2\right)\bar\sigma + \bar\Lambda_0\ell^2\right]
\end{equation}
This equation should apply only when $\gamma\ne0$ (since there is no merger point when $\gamma=0$) and this is the case because 
its derivation assumed $\nu^2\ne1$.

%%%%%%%%%%%%%%%%%%%%%%%%%%%%%%%%%%
%%%%%%%%%%%%%%%%%%%%%%%%%%%%%%%%%
\section{The MMG source tensor}\label{sec:source}

Although the MMG equation (\ref{sourcefree}) is not the result of varying any action for the metric alone, there is an action involving auxiliary fields, given in \cite{Bergshoeff:2014pca}, for which the joint equations for the metric and auxiliary fields are equivalent to the MMG equation for the metric alone; these equations both imply (\ref{sourcefree}) and determine the auxiliary fields in terms of the metric. The unusual feature of this action is that the auxiliary fields are not determined by the auxiliary field equations alone; it is also necessary to use the equation of motion that follows from variation of the metric (more precisely, the dreibein). As a consequence of this feature, back-substitution of the expressions for the auxiliary fields into the {\it action} is not legitimate; this step would produce a new action for the metric alone but the variation of this new action would not then yield (\ref{sourcefree}). This much was already explained in \cite{Bergshoeff:2014pca}.  

The MMG action with auxiliary fields is given in \cite{Bergshoeff:2014pca} in terms of a dreibein 1-form $e^a$ rather than a metric; the auxiliary fields are an independent dualized spin connection 1-form 
$\omega^a$ and another auxiliary Lorentz-vector valued 1-form $h^a$. The action is proportional to the integral of the following Lagrangian 3-form 
\begin{equation}\label{Lag}
L_{MMG} = -\sigma e_a R^a(\omega) + \frac{\Lambda_0}{6}\epsilon_{abc}e^ae^be^c + h_a D(\omega)e^a + \frac{\alpha}{2}\epsilon_{abc}e^ah^bh^c + \frac{1}{\mu}L_{LCS}\, , 
\end{equation}
where  $R^a(\omega)$ is the dualized curvature 2-form, $D(\omega)$ the exterior covariant derivative and $L_{LCS}$  the parity-violating Lorentz-Chern-Simons term, all defined in terms of  the independent spin  connection 1-form $\omega^a$ (we refer the reader to \cite{Bergshoeff:2014pca} for details of the conventions).  Both $\sigma$ and $\alpha$ are dimensionless parameters; it was assumed in \cite{Bergshoeff:2014pca} that $\sigma^2=1$ but {\it here we drop this assumption}.

We assume invertibility of the dreibein as this is needed to construct the metric, and then to show that it satisfies the equation (\ref{sourcefree}). Given this metric we can couple it minimally to matter. If $I_M$ is the resulting generally covariant matter action; then the symmetric (Belinfante) stress-tensor for matter in an arbitrary metric background is 
\begin{equation}
T_{\mu\nu} = - \frac{2}{\sqrt{-\det g}} \, \frac{\delta I_M}{\delta g^{\mu\nu}}\, . 
\end{equation}
It is important to appreciate here that if an affine connection is needed to construct $I_M$ then this must be the torsion-free Levi-Civita connection; in other words, we construct $I_M$ so that it depends on $e^a$ (through the metric) but not on $\omega^a$ or $h^a$. Since  $I_{\rm mat}$ is diffeomorphism invariant, it then follows that the matter field equations imply the condition (\ref{Tcon}).  By this construction, the matter Lagrangian density is the dual of a  3-form $L_M$ with the following variation with respect to a variation of the dreibein
\begin{equation}
\delta L_M =  e\, T^{\mu}{}_{a} \, \delta e_{\mu}{}^a \, , \qquad T^\mu{}_a = T^{\mu\nu} e_{\nu\, a}\, , 
\end{equation}
where $e=\det e_\mu{}^a$. 

We now consider the action matter-coupled MMG action 
\begin{equation}
I= I_{MMG} + I_M\, .
\end{equation}
We  aim to eliminate the auxiliary fields appearing in $I_{MMG}$ to get a modification of the metric equation (\ref{sourcefree}) that involves the matter stress tensor.   We shall now spell out the details of this computation. Following \cite{Bergshoeff:2014pca} we define the new spin-connection
\begin{equation}
\Omega^a = \omega^a + \alpha h^a\, . 
\end{equation}
The $h^a$ field equation still implies that $\Omega$ is torsion-free and hence equal to the standard spin-connection constructed from the dreibein. 
An appropriate linear combination of the $\omega^a$ and $e^a$ field equations then gives
\begin{equation}\label{lincom}
0 = \frac12 \epsilon^{\mu\nu\rho}R_{\nu\rho}{}^a(\Omega) + \mu(1+\sigma\alpha)^2\epsilon^{\mu\nu\rho}\epsilon^{abc}e_{\nu b}h_{\rho c} + \frac{\alpha \Lambda_0}{2}\epsilon^{\mu\nu\rho}\epsilon^{abc}e_{\nu b}e_{\rho c} + \alpha e \, T^{\mu a}\,. 
\end{equation}
Here we use  $\epsilon^{\mu\nu\rho}$ to denote the tensor   $e^{-1}\varepsilon^{\mu\nu\rho}$, where $\varepsilon^{\mu\nu\rho}$ is the invariant alternating tensor density; it then follows that 
\begin{equation}
\epsilon^{abc} = \epsilon^{\mu\nu\rho} e_\mu{}^a e_\nu{}^b e_\rho{}^c\, . 
\end{equation}
Equation (\ref{lincom}) can be solved for $h^a$ or, equivalently for $h_{\mu\nu}$; the solution is
\begin{equation}\label{hsoln}
h_{\mu\nu} = -\frac{1}{\mu (1+\sigma\alpha)^2} \left[ S_{\mu \nu} + \frac{\alpha \Lambda_0}{2} g_{\mu\nu} + \alpha (T_{\mu\nu} - \frac12 T g_{\mu \nu} ) \right] \,, 
\end{equation}
where $S_{\mu\nu}$ is the Schouten tensor; as this is symmetric when defined in terms of the usual torsion-free connection, we deduce that $h_{\mu\nu}$ is symmetric.

The other independent linear combination of the $\omega^a$ and $e^a$ equations of motion is
\begin{eqnarray}\label{pre-eq}
0 &=& \epsilon^{\mu\nu\rho}D_{\nu}(\Omega)h_{\rho}{}^a - \frac{\alpha}{2}\epsilon^{\mu\nu\rho}\epsilon^{abc}h_{\nu b}h_{\rho c} + \sigma\mu(1 + \sigma\alpha)\epsilon^{\mu\nu\rho}\epsilon^{abc}e_{\nu b}h_{\rho c} \nonumber \\
&& +\,  \frac{\Lambda_0}{2}\epsilon^{\mu\nu\rho}\epsilon^{abc}e_{\nu b}e_{\rho c} + e \, T^{\mu a} \,.
\end{eqnarray}
Substituting the expression for $h_\mu{}^a$ into this equation gives the metric field equation. When $T_{\mu\nu}=0$ we recover the source-free equation (\ref{sourcefree}) with\footnote{These expressions agree with those given in \cite{Bergshoeff:2014pca}  when $\sigma^2=1$, as was assumed there. The expressions given here are valid for arbitrary  $\sigma$.}
\begin{equation}\label{barparam1}
\bar{\sigma} = \sigma\left(1 + \sigma\alpha\right) + \frac{\alpha^2\Lambda_0}{2\mu^2(1+\sigma\alpha)^2} \,, \qquad \gamma = -\frac{\alpha}{(1+\sigma\alpha)^2} 
\end{equation}
and 
\begin{equation}\label{barparam2}
\bar\Lambda_0 = \Lambda_0\left(1 + \sigma\alpha - \frac{\alpha^3\Lambda_0}{4\mu^2(1+\sigma\alpha)^2}\right) \, .
\end{equation}
More generally, we find the equation (\ref{sourced}) with source tensor 
\bea\label{sourcetensor}
{\cal T}_{\mu\nu} &=& \left(1+ \sigma\alpha + \frac{\alpha^2\gamma}{2\mu^2}\Lambda_0\right)T_{\mu\nu} - 
\frac{\alpha}{\mu} \, \epsilon_\mu{}^{\rho\sigma} D_\rho \hat T_{\sigma\nu} \nonumber \\
&& +\ \frac{\alpha\gamma}{\mu^2}\,  \epsilon_\mu{}^{\rho\sigma}\epsilon_\nu{}^{\lambda\kappa} S_{\rho\lambda} \hat T_{\sigma\kappa} + 
\frac{\alpha^2\gamma}{2\mu^2}\, \epsilon_\mu{}^{\rho\sigma} \epsilon_\nu{}^{\lambda\kappa} \hat T_{\rho\lambda}\hat T_{\sigma\kappa}\, , 
\eea
where
\be
\hat T_{\mu\nu} = T_{\mu\nu}- \frac{1}{2}g_{\mu\nu} T\, . 
\ee

A drawback of the above result for the MMG source tensor  is that it depends on the parameters $(\sigma,\alpha,\Lambda_0)$ appearing in the action, rather than on the parameters $(\bar\sigma,\gamma,\bar\Lambda_0)$ that appear directly in the metric field equation. The former are determined in terms of the latter by inversion of  (\ref{barparam1}) and (\ref{barparam2}). Assuming the existence of a $\gamma\to0$ limit,  one can show that 
\begin{eqnarray}\label{perturb}
\alpha &=& - \gamma\left(1+\gamma\bar\sigma\right)^{-2} +  {\cal O}\left(\gamma^3\right) \nonumber \\
\sigma &=& \left(1+\gamma\bar\sigma\right)\bar\sigma   - \frac{1}{2}\gamma^2 \bar\Lambda_0/\mu^2 +  {\cal O}\left(\gamma^3\right) \nonumber \\
\Lambda_0 &=& \left(1+\gamma\bar\sigma\right)\bar\Lambda_0  + {\cal O}\left(\gamma^3\right) \, ,  
\end{eqnarray}
but no exact inversion formula is known. In this sense,  the complete source tensor is still given only implicitly in terms of the parameters of the source-free equation. 
However, the  approximate inversion equations (\ref{perturb}) are exact when $\bar\Lambda_0=0$, and this suggests an alternative approach: we could set $\bar\Lambda_0=0$
and then generate a cosmoslogical term  by including a constant ``dark energy'' component to the stress tensor. This indeed leads to a simpler
formulation of the MMG equation, as we detail in the following section.

%%%%%%%%%%%%%%%%%%%%%%%%%%%%
%%%%%%%%%%%%%%%%%%%%%%%%%
\section{An improved MMG parametrization}

Let us now take the MMG equation to be 
\begin{equation}\label{tildeE}
\frac{1}{\mu} C_{\mu\nu} + \eta G_{\mu\nu} + \frac{\gamma}{\mu^2} J_{\mu\nu} = \Theta_{\mu\nu}(T)\, . 
\end{equation}
where 
\begin{equation}\label{Thetadef}
\Theta_{\mu\nu}(T) = \left. {\cal T}_{\mu\nu}\right|_{\bar\Lambda_0=0, \, \bar\sigma=\eta}
\end{equation}
and we use the $T$-tensor 
\begin{equation}
T_{\mu\nu} = - \bar \rho \, g_{\mu\nu} + \theta_{\mu\nu}\, , 
\ee
for some constant energy density $\bar\rho$; the matter stress tensor is now $\theta_{\mu\nu}$.  On the left hand side of this MMG equation we have replaced $\bar\sigma$ by the new constant $\eta$, for reasons that will become clear, and we have set $\bar\Lambda_0=0$. The idea is that the  background energy density $\bar\rho$ in the source term will be equivalent to a cosmological parameter on the left hand side, and we will confirm this intuition below. 

First, we remark that there are {\it two} values of $\Lambda_0$ that yield $\bar\Lambda_0=0$. From (\ref{barparam2}) we see that these are
\be
{\it either}: \quad \Lambda_0=0\,  \qquad \Rightarrow \quad \bar\sigma = \sigma(1+\sigma\alpha)\, ,  \quad \qquad \qquad \qquad ({\rm case \ 1})
\ee
\be
{\it or} \quad \Lambda_0= -\frac{4(1+\sigma\alpha)}{\alpha^2\gamma} \mu^2 \, 
\quad \Rightarrow \quad \bar\sigma = \frac{1}{\alpha}\left(2+ \sigma\alpha\right)(1+\sigma\alpha)\, ,  \quad ({\rm case \ 2}).
\ee
Case $1$ corresponds to an extension to all orders in $\gamma$ of the perturbative inversion formula of (\ref{perturb}); in this case we may take the $\gamma\to0$ limit to recover
TMG results.  In case $2$ there is no TMG limit. 

Next, we observe that since $\bar\Lambda_0=0$, it follows from  (\ref{barparam1})  (with $\bar\sigma\to\eta$) that 
\be\label{direct}
\gamma = -\frac{\alpha}{(1+\sigma\alpha)^2}\, ,  \qquad 
\eta =  \left\{ \begin{array}{ccc}  \sigma(1+\sigma\alpha)  & & ({\rm case  \ 1}) \\  \sigma(1+ \alpha\sigma) + \frac{2}{\alpha}(1+\alpha\sigma)& & ({\rm case  \ 2}) \end{array}\right.
\ee
The advantage of the new form of  the MMG equation is that we have an explicit inversion of these relations, {\it and} we have it for either case:
\be
\alpha= - \frac{\gamma}{(1+\gamma\eta)^2} \, , \qquad 
\sigma= \left\{ \begin{array}{ccc} \eta(1+\gamma\eta)  & & ({\rm case  \ 1}) \\  \eta(1+\gamma\eta) + \frac{2}{\gamma} (1+\gamma\eta) & & ({\rm case  \ 2}) \end{array}\right.
\ee
Notice that the inverse map from $(\eta,\gamma)$ to $(\bar\sigma,\alpha)$ is {\it identical} to the map from $(\bar\sigma,\alpha)$ to $(\eta,\gamma)$, which suggests that some deeper understanding of it should be possible.  Notice too that
\be
(1+ \sigma\alpha)= \left\{ \begin{array}{ccc} (1+\gamma\eta)^{-1} & &  ({\rm case \ 1}) \\ -  (1+\gamma\eta)^{-1} & &  ({\rm case \ 2})\end{array}\right.
\ee
We can now eliminate the parameters $(\sigma,\alpha)$ from the source tensor  in terms of the parameters $(\eta,\gamma)$ of the source-free  MMG equation. Since $\bar\Lambda_0=0$ now, we find {\it for either case} that 
\bea\label{newsource}
 \Theta_{\mu\nu}(T) &=& \frac{1}{(1+\gamma\eta)} T_{\mu\nu} + \frac{\gamma}{\mu(1+\gamma\eta)^2}\epsilon_\mu{}^{\rho\sigma} D_\rho \hat T_{\sigma\nu} 
- \frac{\gamma^2}{(1+\gamma\eta)^2\mu^2} \epsilon_\mu{}^{\rho\sigma} \epsilon_\nu{}^{\lambda\kappa} S_{\rho\lambda}\hat T_{\sigma\kappa} \nonumber \\
&&+\  \frac{\gamma^3}{2(1+\gamma\eta)^4 \mu^2}  \epsilon_\mu{}^{\rho\sigma} \epsilon_\nu{}^{\lambda\kappa} \hat T_{\rho\lambda}\hat T_{\sigma\kappa}\, . 
\eea
We now have an MMG equation with a source tensor $\Theta$ expressed in terms of the dimensionless parameters $(\eta,\gamma)$ of the source-free equation and a background energy density parameter $\bar\rho$. 

Now we shall show that this reformulation of the MMG equation is equivalent to the original formulation in which $\bar\Lambda_0$ may be non-zero and the source tensor
is the ${\cal T}$-tensor as given in (\ref{sourcetensor}) . We first rewrite the equation for $\Theta_{\mu\nu}$ by decomposing the $T$-tensor as in (\ref{Thetadef}).  Using  the identity
\be
\epsilon_\mu{}^{\rho\sigma}\epsilon_\nu{}^{\lambda\kappa} g_{\sigma\kappa} \equiv \delta_\mu^\lambda\delta_\nu^\rho - g_{\mu\nu}g^{\rho\lambda}\, , 
\ee
one finds that 
\be
 \Theta_{\mu\nu}(T)=  -\frac{\bar\rho}{(1+ \gamma\eta)} \left[ 1 + \frac{\bar\rho\gamma^3}{4\mu^2(1+\gamma\eta)^3} \right] g_{\mu\nu} - \frac{\bar\rho\gamma^2}{2\mu^2(1+\gamma\eta)^2} G_{\mu\nu}
+ {\cal T}_{\mu\nu}(\theta)
\ee
where\footnote{The hat notation has the same meaning as before, i.e. $\hat\theta_{\mu\nu} = \theta_{\mu\nu} - (1/2)g_{\mu\nu} \theta$.}
\bea
{\cal T}_{\mu\nu}(\theta) &=& \Theta_{\mu\nu}(\theta)  + \frac{\bar\rho\gamma^3}{2\mu^2(1+\gamma\eta)^4} \theta_{\mu\nu}\nonumber \\
&=& \frac{1}{(1+\gamma\eta)} \left[1+ \frac{\bar\rho\gamma^3}{2\mu^2(1+\gamma\eta)^3}\right]\theta_{\mu\nu} 
+ \frac{\gamma}{\mu(1+\gamma\eta)^2}\epsilon_\mu{}^{\rho\sigma} D_\rho \hat \theta_{\sigma\nu} \\
&& -\ \frac{\gamma^2}{\mu^2(1+\gamma\eta)^2} \epsilon_\mu{}^{\rho\sigma} \epsilon_\nu{}^{\lambda\kappa} S_{\rho\lambda}\hat\theta_{\sigma\kappa} 
+  \frac{\gamma^3}{2\mu^2(1+\gamma\eta)^4}  \epsilon_\mu{}^{\rho\sigma} \epsilon_\nu{}^{\lambda\kappa} \hat\theta_{\rho\lambda}\hat\theta_{\sigma\kappa}\, . 
\nonumber 
\eea
Taking over to the left hand side of the MMG equation the terms in $\Theta_{\mu\nu}(T)$ proportional to the metric and the Einstein tensor, we recover the $E$-tensor   of 
(\ref{sourcefree}) with  
\be\label{direct2}
\bar\sigma = \eta + \frac{\bar\rho\gamma^2}{2\mu^2(1+\gamma\eta)^2}\, , \qquad \bar\Lambda_0 = \frac{\bar\rho}{(1+\gamma\eta)} \left[1+ \frac{\bar\rho\gamma^3}{4\mu^2(1+\gamma\eta)^3}\right]\, . 
\ee
So a cosmological term is indeed generated by a ``dark energy'' component to the $T$-tensor, but with a renormalization of the Einstein tensor coefficient. 
If we now choose
\be\label{choosea}
\bar\rho= \left\{ \begin{array}{ccc}  \Lambda_0 & & ({\rm case \ 1})
\\  \Lambda_0 - \frac{4}{\gamma^3} (1+\gamma\eta)^3\mu^2 & & ({\rm case \ 2})\end{array} \right.
\ee
and use (\ref{direct}) for $\eta$,  we recover the expression of (\ref{barparam1}) for $\bar\sigma$, and the expression (\ref{barparam2}) for $\bar\Lambda_0$. Furthermore, for this choice of $\bar\rho$ the tensor   ${\cal T}_{\mu\nu}(\theta)$ is, {\it for either case}, precisley  the original source tensor  of (\ref{sourcetensor}) but with the renamed matter 
stress tensor. 

Finally, we present some details of an explicit check of the consistency of the MMG equation with  source tensor. Recall that the requirement for consistency is that 
the source tensor ${\cal T}$ satisfy (\ref{identity}). For the reformulation of the MMG equation in which any explicit cosmological term is absent this consistency condition reads 
\begin{equation}
D_\mu\Theta^{\mu\nu}(T) = \frac{\gamma}{\mu} \epsilon^{\nu\rho\sigma} S_\rho{}^\tau \Theta_{\sigma\tau}(T)\, . 
\end{equation}
Using the expression (\ref{newsource}) for the $\Theta$-tensor we find that 
\bea
LHS &=& \frac{\gamma}{2\mu(1+\gamma\eta)^2} \, \epsilon^{\mu\rho\sigma} [D_\mu,D_\rho] \hat T_{\sigma\nu} - \frac{\gamma^2}{\mu^2(1+\gamma\eta)^2} \epsilon_\nu{}^{\lambda\kappa}C_{\lambda\sigma} \hat T^\sigma{}_\kappa 
\nonumber \\
&& +\ \frac{\gamma^2}{\mu^2(1+\gamma\eta)^2} \, \epsilon_\nu{}^{\lambda\kappa} \left[S_{\rho\lambda} - \frac{\gamma}{(1+\gamma\eta)^2} \hat T_{\rho\lambda}\right] \left(\epsilon^{\rho\mu\sigma}D_\mu\hat T_{\sigma\kappa} \right) \, . 
\eea
Now we use, in the first term\footnote{This makes use of the 3D identity $R_{\mu\nu\rho\sigma} = g_{\mu \rho}S_{\nu \sigma} - g_{\nu \rho}S_{\mu \sigma} - g_{\mu \sigma}S_{\nu \rho} + g_{\nu \sigma}S_{\mu \rho}$.}
\be
\epsilon^{\mu\rho\sigma} [D_\mu,D_\rho] \hat T_{\sigma\nu}  = 2\epsilon_\nu{}^{\rho\sigma}S_{\rho\lambda} \hat T^\lambda{}_\sigma\, . 
\ee
In the second term we use the MMG equation in the form of  (\ref{tildeE}) to eliminate $C$. Freely using  the identity
\be
\epsilon^{\rho\mu\sigma}D_\mu\hat T_{\sigma\kappa} \equiv \epsilon_\kappa{}^{\mu\sigma}D_\mu \hat T_\sigma{}^\rho\, , 
\ee
we then find that 
\bea
LHS &=& - \frac{\gamma}{\mu} \epsilon_\nu{}^{\sigma\rho} S_\rho{}^\lambda\left[ \frac{1}{(1+\gamma\eta} T_{\sigma\lambda} + \frac{\gamma}{\mu(1+\gamma\eta)^2} \epsilon_\lambda{}^{\mu\eta}D_\mu\hat T_{\eta\sigma} \right] \\
&&+ \ \frac{\gamma^3}{\mu^3(1+\gamma\eta)^2} \epsilon_\nu{}^{\lambda\kappa} J_{\lambda\sigma}\hat T^\sigma{}_\kappa 
+ \frac{\gamma^4}{\mu^2(1+\gamma\eta)^4} \epsilon_\nu{}^{\lambda\kappa}\epsilon_\lambda{}^{\alpha\beta} \epsilon_\sigma{}^{\gamma\delta} S_{\alpha\gamma}  \hat T^\sigma{}_\kappa \, . \nonumber 
\eea
Here we have used the fact that some terms with a factor of $\gamma^3$ in the numerator cancel, and that the term with a factor of $\gamma^5$ in the numerator is identically zero. Next, we add and subtract terms in the first line, and expand the term involving $J$ in the second, to get
\bea
LHS &=& - \frac{\gamma}{\mu} \epsilon_\nu{}^{\sigma\rho}S_\rho{}^\lambda \Theta_{\sigma\lambda}(T)  -  \frac{\gamma^3}{\mu^2(1+\gamma\eta)^2} \epsilon_\nu{}^{\sigma\rho} \epsilon_\sigma{}^{\alpha\beta}\epsilon^{\lambda\gamma\delta} S_{\alpha\beta} 
\left[S_{\rho\lambda}\hat T_{\beta\delta} + \frac{1}{2}S_{\beta\delta} \hat T_{\rho\lambda}\right] \nonumber \\
&& + \frac{\gamma^4}{\mu^2(1+\gamma\eta)^4} \epsilon_\nu{}^{\sigma\rho}\epsilon_\sigma{}^{\alpha\beta} \epsilon^{\lambda\gamma\delta} 
\hat T_{\beta\delta}\left[ S_{\alpha\gamma}\hat T_{\rho\lambda} + \frac{1}{2}S_{\rho\lambda}\hat T_{\alpha\gamma}\right]\, . 
\eea
The second and third terms are identically zero, so we have
\be
LHS= - \frac{\gamma}{\mu} \epsilon_\nu{}^{\sigma\rho}S_\rho{}^\lambda \Theta_{\sigma\lambda}(T)  =RHS\, . 
\ee

%%%%%%%%%%%%%%%%%%%%%%%%%%%%%%%%%%%%%
%%%%%%%%%%%%%%%%%%%%%%%%%%%%%%%%%%%%
\section{Ideal fluid cosmologies}

We will now look for solutions to the matter-coupled MMG equation  for which the metric takes the standard FLRW form
\begin{equation}
ds^2 = -dt^2 + a^2(t)\left(\frac{dr^2}{1-kr^2} + r^2d\theta^2 \right)\, , 
\end{equation}
where we may restrict to $k=-1,0,1$ without loss of generality, corresponding to open, flat and closed 3D universes. 
For an ideal fluid source the  $tt$ component of the MMG equation is 
\begin{equation}\label{tteq}
\gamma \left[\left(\frac{\dot a}{a}\right)^2 + \frac{k}{a^2} + \alpha\rho\right]^2 -4\mu^2\bar\sigma \left[ \left(\frac{\dot a}{a}\right)^2 + \frac{k}{a^2} + \alpha\rho\right]
= 4\mu^2\left( \frac{\alpha\rho}{\gamma} - \bar\Lambda_0\right)\, . 
\end{equation}
The remaining components of the MMG equation are then satisfied identically provided that the matter stress tensor satisfies $D^\mu T_{\mu\nu}=0$; for an ideal fluid this is equivalent to the continuity equation
\begin{equation}
\dot\rho = - 2\left(\frac{\dot a}{a}\right) \left(\rho+p\right)\, . 
\end{equation}

Equation (\ref{tteq}) can be rewritten in the Friedmann equation form 
\begin{equation}\label{Fried}
\left(\frac{\dot a}{a}\right)^2 + \frac{k}{a^2} = \Lambda+ \rho_{\rm eff}\, ,  \qquad \Lambda\equiv \frac{2\mu}{\gamma}\left( \mu\bar\sigma -m\right)
\end{equation}
where $m$ is the mass parameter defined in (\ref{mp}), and the  ``effective'' energy density is 
\begin{equation}\label{rhoeff}
\rho_{\rm eff} =- \frac{2\mu}{\gamma}\left[ \sqrt{\alpha\rho + m^2} - m\right] - \alpha\rho\, . 
\end{equation}
There are sign ambiguities inherent in the square roots. Here we choose the signs such  that the $\gamma\to0$ limit exists,  and such that $\rho_{\rm eff} \to \rho$ in this limit. 
Notice that (\ref{Fried}) is {\it exactly} the Friedmann equation in the source-free  $(\rho=0$) case; the MMG parameter $\gamma$
appears only in the equation relating the cosmological constant $\Lambda$ to the parameters of the model.  Thus, source-free MMG cosmologies are precisely
those of TMG. However, the two models differ dramatically when a source is included, as we shall see. 

If we assume a linear equation of state of the form\footnote{$\kappa=-1$ is equivalent to a cosmological constant, which is already included.}
\begin{equation}
p(t)= \kappa \rho(t)\, , \qquad  -1<\kappa \le1\, , 
\end{equation}
then the  continuity equation becomes, for $\rho\ne0$, 
\begin{equation}\label{cont}
2\left(1+\kappa\right) \left(\frac{\dot a}{a}\right) + \left(\frac{\dot\rho}{\rho}\right) =0 \, , 
\end{equation}
which implies that 
\begin{equation}
\rho= \frac{\rho_0}{a^{2(1+\kappa)}}\, 
\end{equation}
for some constant (co-moving) energy density $\rho_0$. 

Let us begin by supposing that $\rho$ is small at late times; in this case, we have 
\begin{equation}
\rho_{\rm eff}= \left(1+\sigma\alpha\right)^2 \left[1+ \gamma\,  m/\mu\right] \rho + {\cal O}\left(\rho^2\right)
\end{equation}
at late times. Apart from the overall factor multiplying $\rho$, this leads to the standard Friedmann equation, in which the density is negligible at late times, implying
an expanding, asymptotically dS,  universe as $|t|\to \infty$. Let us consider such a universe near $t=\infty$ and run the expansion backwards in time: as the universe contracts
the density increases.  For simplicity we now restrict to the $k=0$ case, for which the equation to solve for $\rho(t)$ is 
\begin{equation}\label{tosolve}
\left(\frac{\dot\rho}{\rho}\right)^2 = 4\left(1+\kappa\right)^2 \left[\Lambda+ \rho_{\rm eff}\right]\, . 
\end{equation}
This equation requires the ``total'' energy density $\Lambda+\rho_{\rm eff}$ to be positive, but we can rewrite this as follows:
\begin{equation}
\Lambda+\rho_{\rm eff} = \left[ \Lambda+ \left(m + \frac{\mu}{\gamma}\right)^2\right] - \xi^2\, , \qquad \xi = \sqrt{\alpha\rho + m^2} + \frac{\mu}{\gamma}\, . 
\end{equation}
Positivity of this expression implies an upper bound on the variable $\xi$. If $\alpha>0$ this upper bound implies an upper bound on $\rho$.  However, if $\alpha<0$  then there is already an upper bound on $\rho$ due to the requirement that $\alpha\rho + m^2$ be non-negative, so we must consider separately the cases of positive and negative $\alpha$.

%%%%%%%%%%%%%%%%%%%%%%%%%%%%
\subsection{Positive $\alpha$}

In this case we have 
\begin{equation}
\xi_{\rm max} = \sqrt{\alpha\rho_{\rm max} +m^2} + \frac{\mu}{\gamma}\, , \qquad 
\xi_{\rm max} =  \sqrt{\Lambda +  \left(m + \frac{\mu}{\gamma}\right)^2}\, . 
\end{equation}
Near the maximum density we can write
\begin{equation}
\rho(t) = \rho_{\rm max} - \delta(t)\, , 
\end{equation}
in which case 
\begin{equation}
\Lambda + \rho_{\rm eff} = \left(\frac{\xi_{\max}}{\xi_{\rm max} - \mu/\gamma}\right) \delta + {\cal O}\left(\delta^2\right)\, . 
\end{equation}
Substituting this into (\ref{tosolve}) and ignoring non-linear terms in $\delta$ we find that 
\begin{equation}
\dot \delta ^2 \propto \delta\, , 
\end{equation}
which implies that $\delta \sim t^2$ if we choose the time coordinate such that $t=0$ when $\rho=\rho_{\rm max}$. Notice that this solution is invariant under time reversal $t\to -t$, as is the full equation (\ref{tosolve}). This means that as the universe contracts (running it backwards in time from $t=\infty$) the density passes smoothly through a maximum value
at which the scale factor takes its minimum value, after which the universe re-expands (now running backwards in time from $t=0$ to $t=-\infty$). 

In the special case that $m=0$, which is the merger point, it is not difficult to find the explicit solution. The equation to solve in this case is
\begin{equation}\label{xeq}
\left(\frac{\dot x }{x}\right)^2 = -\left(x-a_+\right)\left(x-a_-\right)\, , \qquad a_\pm = \frac{\mu\left(1+\kappa\right)}{\gamma}  \left[1\pm \sqrt{1+ \gamma^2\Lambda/\mu^2}\right] \, ,  
\end{equation}
where 
\begin{equation}
x= \left(1+\kappa\right) \sqrt{\alpha\rho} \, , \qquad x>0\, . 
\end{equation}
The restriction to positive $x$ follows from the choice of signs of square roots that leads to (\ref{Fried}). 

As $\gamma$ is negative, we have $a_+<0$ and $a_->0$ when $\mu>0$.  In this case we have the further restriction  $x\le a_-$. Taking the square root on both sides of (\ref{xeq}) then yields 
\begin{equation}\label{xdot}
\dot x= \pm x \sqrt{\left(a_- -x\right) \left(x-a_+\right)}\, . 
\end{equation}
Since $x>0$ there is a one fixed point solution with $x= a_-$; this corresponds to an  unstable Minkowski vacua with a constant energy density such that 
$\Lambda + \rho_{\rm eff}=0$. There is also the time-dependent solution 
\begin{equation}
x(t) =  \frac{-\gamma (1+\kappa)\Lambda}{\sqrt{\gamma^2\Lambda + \mu^2}\, \cosh\left[\sqrt{\Lambda}(1+\kappa)t\right] + \mu}\ . 
\end{equation}
This is a time-symmetric solution that approaches the dS vacuum as  $|t|\to\infty$, and passes smoothly through the unstable Minkowski vacuum at $t=0$. The correspomding solution for the scale factor is 
\begin{equation}
a(t)^{\kappa +1}=\frac{\sqrt{\alpha \rho_0}}{{|\gamma| \Lambda}} \left(\sqrt{\gamma^2 \Lambda +\mu^2} \cosh\left(\sqrt{\Lambda} (1+\kappa) t\right) + \mu  \right) \, , 
\end{equation}
which shows that $a(t)$ passes smoothly through a minimum value at $t=0$. 

Changing the sign of $\mu$ (so that $\mu<0$) flips the signs of $a_\pm$, so now $a_+>0$ and $a_-<0$. Then (\ref{xdot}) is replaced by 
\begin{equation}
\dot x= \pm x \sqrt{\left(a_+ -x\right) \left(x-a_-\right)}\, .
\end{equation}
and we have same solution as before, but with $\mu\to-\mu$. Notice that changing the sign of $\mu$, so that $\mu<0$,  changes the solution since the maximum density, reached at $t=0$, is now larger. It would appear from this fact that the sign of $\mu$ is physically relevant but here we should recall that we made a choice of sign for the square root  in passing from  (\ref{tteq}) to (\ref{Fried}), and a change in this sign will cancel the effect of a change of sign of  $\mu$. So the two distinct $m=0$ cosmologies are both possible for either sign of $\mu$.

%%%%%%%%%%%%%%%%%%%%%%%%%
\subsection{Negative $\alpha$} 

When $\alpha<0$ we have the bound $\rho \le \rho_{\rm max}$, where now $\rho_{\rm max} = m^2/|\alpha|$. 
At this maximum density we can again write
\begin{equation}
\rho(t) = \rho_{\rm max}- \delta(t)\, , 
\end{equation}
in which case
\begin{equation}
\Lambda+ \rho_{\rm eff} = \Lambda^{\rm (total)}_{\rm max} - \frac{\mu \sqrt{|\alpha|}}{\gamma} \sqrt{\delta} + {\cal O}(\delta)\, , 
\end{equation}
where
\begin{equation}
\Lambda^{\rm (total)}_{\rm max} = \Lambda + \left(m+\frac{\mu}{\gamma}\right)^2 - \frac{\mu^2}{\gamma^2}\, .
\end{equation}
Since $m$ is assumed positive and $\gamma$ is positive when $\alpha<0$, we see that $\Lambda^{\rm (total)}_{\rm max} >0$ when $\Lambda>0$, as 
we also assume here. This means that (\ref{tosolve}) now takes the following form for small $\delta$:
\begin{equation}
\dot\delta^2 = c^2 + {\cal O}\left( \sqrt{\delta}\right) \, , 
\end{equation}
for some non-zero constant $c$. This implies that $\delta \sim t$ near $t=0$ (assuming again that we choose the time coordinate such that $t=0$ at the maximum density) so that  
$\delta$ becomes negative for small negative $t$, violating the upper bound on $\rho$. There is therefore a singularity at $t=0$, but it is not a  big-bang singularity  because 
$\rho$ is finite and $a\ne0$ at $t=0$. 

This singular behaviour suggests that  the model is unphysical when $\alpha<0$, so we must assume that $\alpha>0$, and hence that $\gamma<0$.

%%%%%%%%%%%%%%%%%%%%%%%%%%%%%%%%%%%%
%%%%%%%%%%%%%%%%%%%%%%%%%%%%%%%%%%%%
\section{Discussion} 

The recent 3D ``minimal massive gravity'' (MMG) model \cite{Bergshoeff:2014pca} resolves a number of the difficulties of the much-studied ``topologically massive gravity'' (TMG) model 
in the context of a possible holographic dual CFT in asymptotically anti-de Sitter spacetimes,  but it has very special properties that seem to rule out coupling to matter;  this was one reason for the 
epithet ``minimal'', the other being that, like TMG, its only local degree of freedom is a graviton with  one polarisation state. In this paper we have shown that matter coupling is possible
but it requires a source tensor ${\cal T}$ that replaces the usual stress tensor $T$ in the coupling of matter to TMG. We have constructed this source tensor explicitly; it turns out to be quadratic in the matter stress tensor. 

The construction relies of the existence of an MMG action with auxiliary fields (which can be eliminated from the resulting equations but not from the action itself). As a result, the 
source tensor is found in terms of the parameters of this action rather than in terms of the parameters in the source-free equation. The latter are definite functions of the former but
the inverse relation is needed to find  the source tensor in terms of the parameters of the source-free equation, and no explicit inversion formula is known.  However, our construction allowed us to circumvent this problem by moving the cosmological term from the source-free part of the MMG equation to a dark energy component of the source. This leads to a reparametrization of the the MMG equations in terms of two dimensionless parameters  of the source-free equation and the constant background energy density, and in terms of these parameters we have found a closed-form expression for  the source-tensor. 

We have explored some of the consequences of this result in the context of cosmological solutions driven by a positive cosmological constant and an ideal fluid source with a linear equation of state.  For TMG one finds the usual Friedmann equation for the scale factor of the (3D) universe, with the usual relation to the energy density. In the MMG case the energy density gets replaced by an ``effective energy density'', which alters the solutions dramatically  as the universe contracts;  the usual big-bang singularity is replaced (at least for flat universes) by a smooth bounce connecting two isometric asymptotic de Sitter spacetimes. 
Of course, this has no obvious consequences for 4D cosmology but it does illustrate how MMG has better short-distance behaviour than TMG. 

These cosmological results apply on the assumption that the sign of the parameter $\gamma$ in the MMG equation is negative, since positive $\gamma$ leads to unacceptable singularities at non-zero scale factor. The sign of $\gamma$ is not actually physically significant by itself because a change of sign in the MMG equation can be compensated by a change in sign of the mass parameter $\mu$ and the other source-free parameters and a change in sign of the source tensor, but  we have effectively removed this sign ambiguity
by a choice of sign for the action. The same sign choice was made in   \cite{Bergshoeff:2014pca}, where it was then found that $\gamma>0$ is required for unitarity in AdS. This does not contradict the results found here, of course, because the dS and AdS vacua correspond to  disjoint regions of  parameter space. However, it would be interesting to see what the analogous implications are for domain wall solutions asymptotic to an AdS vacuum, which may be related to the cosmological solutions considered here by some extension of  the cosmology/domain-wall correspondence \cite{Skenderis:2006fb}. 

A novel feature of MMG, as compared to TMG, is that the cosmological constant $\Lambda$ of a vacuum solution of the MMG equation is determined by a quadratic equation, so that there are generically either two possible maximally-symmetric vacua for a given value of the cosmological parameter  appearing in the MMG equation, or none.  In this respect, MMG  is similar to the parity-preserving ``new massive gravity''  model. As in that case, the ``merger point'' at which the two distinct maximally-symmetric vacua coincide has
special properties; in  particular, we have shown here, for $\Lambda\ne0$, that  static black hole solutions exist that are asymptotic but not locally isometric to the unique (A)dS$_3$
vacuum. In the extremal limit these solutions interpolate between the (A)dS$_3$ vacuum and a Kaluza-Klein  (A)dS$_2 \times S^1$ vacuum.  

We have also shown that MMG admits ``warped (A)dS$_3$'' vacua, which raises the question as to whether MMG still resolves the ``bulk vs boundary clash'' of TMG for 
non-zero warp factor. We leave this and other ``warped'' issues to future research. 
\vfill\eject

\section*{Acknowledgements}
A.S.A. and A.J.R. acknowledge support from the UK Science and Technology Facilities Council. A.S.A also acknowledges support from Clare Hall College, Cambridge, and the Cambridge Trust.

%\bibliography{refsspin}

\begin{thebibliography}{1}

%\cite{Deser:1981wh}
\bibitem{Deser:1981wh}
S.~Deser, R.~Jackiw, and S.~Templeton, ``{Topologically Massive Gauge
  Theories},'' {\em Annals Phys.} {\bf 140} (1982)
372--411.
%%CITATION = APNYA,140,372;%%.

%\cite{Brown:1986nw}
\bibitem{Brown:1986nw}
  J.~D.~Brown and M.~Henneaux,
  ``Central Charges in the Canonical Realization of Asymptotic Symmetries: An Example from Three-Dimensional Gravity,''
{\em   Commun.\ Math.\ Phys}.\  {\bf 104} (1986) 207.
  %%CITATION = CMPHA,104,207;%%
  
  %\cite{Kraus:2005zm}
\bibitem{Kraus:2005zm}
  P.~Kraus and F.~Larsen,
  ``Holographic gravitational anomalies,''
  JHEP {\bf 0601} (2006) 022
  [hep-th/0508218].
  %%CITATION = HEP-TH/0508218;%%
  
  %\cite{Kraus:2006wn}
\bibitem{Kraus:2006wn}
  P.~Kraus,
  ``Lectures on black holes and the AdS(3) / CFT(2) correspondence,''
  Lect.\ Notes Phys.\  {\bf 755} (2008) 193
  [hep-th/0609074].
  %%CITATION = HEP-TH/0609074;%%


  
  %\cite{Bergshoeff:2014pca}
\bibitem{Bergshoeff:2014pca}
  E.~Bergshoeff, O.~Hohm, W.~Merbis, A.~J.~Routh and P.~K.~Townsend,
  ``Minimal Massive 3D Gravity,'' Class. Quantum Grav. 31 (2014) 145008
 [ arXiv:1404.2867 [hep-th]].
  %%CITATION = ARXIV:1404.2867;%%
  
  %\cite{Bergshoeff:2009hq}
\bibitem{Bergshoeff:2009hq}
  E.~A.~Bergshoeff, O.~Hohm and P.~K.~Townsend,
  ``Massive Gravity in Three Dimensions,''
  Phys.\ Rev.\ Lett.\  {\bf 102} (2009) 201301
  [arXiv:0901.1766 [hep-th]]; 
  %%CITATION = ARXIV:0901.1766;%%
  
  
  %\cite{Bergshoeff:2009aq}
\bibitem{Bergshoeff:2009aq}
  E.~A.~Bergshoeff, O.~Hohm and P.~K.~Townsend,
  ``More on Massive 3D Gravity,''
  Phys.\ Rev.\ D {\bf 79} (2009) 124042
  [arXiv:0905.1259 [hep-th]].
  %%CITATION = ARXIV:0905.1259;%%
  
   %\cite{Clement:2009gq}
\bibitem{Clement:2009gq}
  G.~Clement,
  ``Warped AdS(3) black holes in new massive gravity,''
  Class.\ Quant.\ Grav.\  {\bf 26} (2009) 105015
  [arXiv:0902.4634 [hep-th]].
  %%CITATION = ARXIV:0902.4634;%%

  
%\cite{Nutku:1993eb}
\bibitem{Nutku:1993eb}
  Y.~Nutku,
  ``Exact solutions of topologically massive gravity with a cosmological constant,''
  Class.\ Quant.\ Grav.\  {\bf 10} (1993) 2657.
  %%CITATION = CQGRD,10,2657;%%
  
  %\cite{Moussa:2003fc}
\bibitem{Moussa:2003fc}
  K.~A.~Moussa, G.~Clement and C.~Leygnac,
  ``The Black holes of topologically massive gravity,''
  Class.\ Quant.\ Grav.\  {\bf 20} (2003) L277
  [gr-qc/0303042].
  %%CITATION = GR-QC/0303042;%%
  
  %\cite{Bouchareb:2007yx}
\bibitem{Bouchareb:2007yx}
  A.~Bouchareb and G.~Clement,
  ``Black hole mass and angular momentum in topologically massive gravity,''
  Class.\ Quant.\ Grav.\  {\bf 24} (2007) 5581
  [arXiv:0706.0263 [gr-qc]].
  %%CITATION = ARXIV:0706.0263;%%
  
 %\cite{Gurses}
\bibitem{Gurses}
 M. G\"urses, 
 ``Perfect fluid sources in 2+ 1 dimensions'', 
 Class. Quantum Grav. {\bf 11}, 2585 (1994)
   
%\cite{Anninos:2008fx}
\bibitem{Anninos:2008fx}
  D.~Anninos, W.~Li, M.~Padi, W.~Song and A.~Strominger,
  ``Warped AdS(3) Black Holes,''
  JHEP {\bf 0903} (2009) 130
  [arXiv:0807.3040 [hep-th]].
  %%CITATION = ARXIV:0807.3040;%%  
   
  
  
   %\cite{Higuchi:1986py}
\bibitem{Higuchi:1986py}
  A.~Higuchi,
  ``Forbidden Mass Range for Spin-2 Field Theory in De Sitter Space-time,''
  Nucl.\ Phys.\ B {\bf 282} (1987) 397.
  %%CITATION = NUPHA,B282,397;%%
  
  %\cite{Gibbons:1993sv}
\bibitem{Gibbons:1993sv}
  G.~W.~Gibbons and P.~K.~Townsend,
  ``Vacuum interpolation in supergravity via super p-branes,''
  Phys.\ Rev.\ Lett.\  {\bf 71} (1993) 3754
  [hep-th/9307049].
  %%CITATION = HEP-TH/9307049;%%
  
 
  
    %\cite{Anninos:2010}
\bibitem{Anninos:2010}
  D.~Anninos,
  ``Sailing from warped AdS3 to warped dS3 in topologically massive gravity,''
  JHEP {\bf 1002} (2010) 046
  [arXiv:0906.1819v2 [hep-th]].
 




%\cite{Skenderis:2006fb}
\bibitem{Skenderis:2006fb}
  K.~Skenderis and P.~K.~Townsend,
  ``Pseudo-Supersymmetry and the Domain-Wall/Cosmology Correspondence,''
  J.\ Phys.\ A {\bf 40} (2007) 6733
  [hep-th/0610253].
  %%CITATION = HEP-TH/0610253;%%
  

\end{thebibliography}
%\bibliographystyle{toine}

\providecommand{\href}[2]{#2}\begingroup\raggedright\endgroup

\end{document}